\documentclass[twocolumn,showpacs,prb,superscriptaddress]{revtex4}

\usepackage{graphicx}
\usepackage{color}
\usepackage{tabularx}
\usepackage{epsfig}
\usepackage{amsmath}
\usepackage{amssymb}
\usepackage{amsfonts}
\usepackage{graphicx}
\usepackage{dcolumn}
\usepackage{bm}
\usepackage{wasysym}
\usepackage{latexsym}
\def\vec{\mathbf}
\usepackage[colorlinks,bookmarks=false,citecolor=blue,linkcolor=red,urlcolor=blue]{hyperref}

\definecolor{grn}{rgb}{0,0,0.54}

\newcommand{\ket}[1]{\ensuremath{| #1 \rangle}}

\begin{document}

\title{Phase separation {\it{versus}} supersolid behavior in frustrated antiferromagnets}


\author{A.~Fabricio Albuquerque}
\affiliation{Laboratoire de Physique Th{\' e}orique, CNRS and Universit{\' e} de
Toulouse, F-31062 Toulouse, France}
\affiliation{School of Physics, The University of New South Wales, Sydney, NSW 2052, Australia}

\author{Nicolas Laflorencie}
\affiliation{Laboratoire de Physique des Solides, Universit\'e Paris-Sud, UMR-8502 CNRS,
91405 Orsay, France}

\author{Jean-David Picon}
\affiliation{Theoretische Physik, ETH Z\"urich, 8093 Z\"urich, Switzerland}
\affiliation{Institute of Theoretical Physics, \'Ecole Polytechnique F\'ed\'erale de Lausanne, CH-1015 Lausanne, Switzerland}

\author{Fr{\' e}d{\' e}ric Mila}
\affiliation{Institute of Theoretical Physics, \'Ecole Polytechnique F\'ed\'erale de Lausanne, CH-1015 Lausanne, Switzerland}


\date{\today}
\pacs{75.10.Jm, 03.75.Nt, 05.30.Jp}

\begin{abstract}
We investigate the competition between spin-supersolidity and phase separation in a frustrated spin-half model of weakly coupled dimers.
We start by considering systems of hard-core bosons on the square lattice, onto which the low-energy physics of the herein investigated
spin model can be mapped, and devise a criterion for gauging the interplay between supersolid order and domain wall formation based
on strong coupling arguments. Effective bosonic models for the spin model are derived via the contractor renormalization (CORE) algorithm
and we propose to combine a self-consistent cluster mean-field solution with our criterion for the occurrence of phase separation to derive
the phase diagram as a function of frustration and magnetic field. In the limit of strong frustration, the model is shown to be unstable
toward phase separation, in contradiction with recently published results. However, a region of stable supersolidity is identified for intermediate
frustration, in a parameter range not investigated so far and of possible experimental relevance.
\end{abstract}

\maketitle


\section{Introduction}
\label{sec:introduction}

Dimer-based antiferromagnets (DAFs) under a magnetic field are promising candidates for displaying new phases of bosonic
matter.\cite{giamarchi:08} Magnetic excitations in such systems, termed {\em triplons}, are well
described by lattice models of interacting bosons, whose density can be finely tuned by varying the magnitude of the
applied field.\cite{Mila98,Totsuka98-Giamarchi99} Experimentally, field-induced Bose-Einstein
condensation (BEC) of triplons has been observed in a number of DAFs (see the review Ref.~\onlinecite{giamarchi:08})
and, remarkably, exotic quantum criticality has been detected in the spin-dimer compound ${\rm BaCuSi_2O_6}$.
\cite{sebastian:06,Ruegg07,Kramer07,Laflorencie09} The presence of magnetic frustration further adds to the rich phenomenology
of these systems by enhancing repulsive interactions between triplons, something that may eventually stabilize
incompressible phases that break the lattice's translational symmetry.\cite{Mila98,Takigawa-Mila10}
Such {\em crystalline} phases are for instance realized in the Shastry-Sutherland material ${\rm
SrCu_{2}(BO_{3})_{2}}$,\cite{kageyama:99,kodama:02} where they are signaled by a series of
magnetization plateaux at unconventional fillings stabilized by complex triplon interactions.
\cite{Dorier08-Abendschein08}

The occurrence of both BEC and solid phases in the phase diagram of DAFs under magnetic field suggests that the
magnetic equivalent of the phase simultaneously displaying diagonal and off-diagonal order known as {\em supersolid}
(SS)\cite{andreev:69,kim:04,Balibar10} may be realized in these systems. Indeed, insofar as more exotic possibilities are
excluded,\cite{senthil:04} according to the Ginzburg-Landau-Wilson paradigm a continuous transition between phases
breaking different symmetries (as it is the case with BEC and crystalline phases) is precluded and we are therefore left
with two possibilities: (i) a first-order transition or (ii) the appearance of an intermediate phase, where both order parameters
coexist, termed spin-supersolid (spin-SS) in the present context. The latter possibility has been first investigated by Momoi
and Totsuka for the Shastry-Sutherland model in the vicinity of half- and third-filling plateaux,\cite{momoi:00} based on a
mean-field analysis of an effective bosonic model derived up to third-order in the inter-dimer coupling. More
recently,\cite{Dorier08-Abendschein08} state-of-the-art techniques have been employed in deriving effective models that
improve upon the third-order effective Hamiltonian of Ref.~\onlinecite{momoi:00}. Unfortunately, the reliability of these methods
is still limited to inter-dimer couplings equal to, at most, one-half of the intra-dimer coupling, and in that parameter range
the different plateaux seem to be separated by first-order transitions without any convincing evidence of spin-SS
phases.\cite{Dorier08-Abendschein08}

The situation is much clearer for the DAF investigated in Ref.~\onlinecite{ng:06-laflorencie:07} where repulsion among
triplons are enhanced due to the strong Ising-like character of the inter-dimer exchange [see Eq.~(1) in Ref.~\onlinecite{picon:08}],
making room for checkerboard solid (CBS) and spin-SS phases to emerge. The absence of frustration allows for quantum
Monte Carlo (QMC) simulations to be performed and, in this way, the occurrence of a spin-SS phase for the model studied in
Ref.~\onlinecite{ng:06-laflorencie:07} has been firmly established. However, such a strongly anisotropic Hamiltonian is unrealistic for
Mott insulating materials and further investigations of models where the kinetic energy is instead reduced by frustration of isotropic
couplings\cite{sengupta:07a,chen:10} are clearly called for if connection to experiments is ever to be made.

In this context, the recent report of a spin-SS phase in a spin-half frustrated DAF by Chen {\em et al.},\cite{chen:10}
who have relied on a novel tensor-product algorithm, is an important result. However, in view of the first-order transitions
observed in the related case of the Shastry-Sutherland model, a systematic investigation of the possibility of phase
separation (PS) is still required.

\begin{figure}
\begin{center}
  \includegraphics*[width=0.18\textwidth,angle=270]{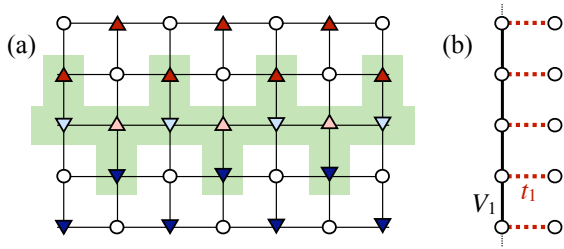}
  \caption{
    (Color online) (a) A DW between mismatching domains in a CBS doped with holes is highlighted: open circles represent
                                    hard-core bosons (triplets) and holes/singlets in the upper (lower) domain are indicated by upward (downward)
                                    triangles; {\em doped} holes are shown as light-filled upward {\em or} downward triangles. (b) A simplified model
                                    for the DW, valid for $V_1/t_1 \gg 1$, is defined on a ``comb" geometry: holes hop (with amplitude $t_1$) through
                                    the links indicated by dashed lines and repel, with strength $V_1$, one another along the vertical nearest-neighbor
                                    links indicated by solid lines.
  }
  \label{fig:DWall}
  \end{center}
\end{figure}

In this paper, we investigate the interplay between SS order and PS in the frustrated DAF analyzed in Ref.~\onlinecite{chen:10}. We begin by
estimating the energetic gains behind PS and supersolidity for hard-core bosons on the square lattice by relying on strong coupling arguments,
and introduce an indicator of the instability toward PS. We then proceed to the analysis of effective bosonic models obtained from the application
of the Contractor Renormalization (CORE) algorithm\cite{morningstar:94-96} to the DAF investigated in Ref.~\onlinecite{chen:10}.
The so-obtained effective Hamiltonians are studied by performing self-consistent cluster mean-field theory (SCMFT) calculations and tendency
toward PS is gauged through means of the aforementioned indicator.


\section{Phase Separation versus Supersolidity for Lattice Bosons}
\label{sec:PS}

In this section, we analyze the interplay between PS and supersolidity in models of hard-core bosons on the square lattice,
onto which the low-energy physics of the spin model considered in the remainder of this paper can be mapped.

\subsection{Instability to Domain Wall Formation}
\label{sec:DW}

We start by considering the simplest model of hard-core bosons on the square lattice, the so-called $t-V$ model:
\begin {equation}
{\mathcal H}_{t-V} = -t_1 \sum_{\left\langle i,j \right\rangle} \left( b^{\dagger}_ib_j + {\rm H.c.}  \right)
                                    +V_1 \sum_{\left\langle i,j \right\rangle}  n_in_j  - \mu \sum_i n_i~.
\label{eq:tVmodel}
\end {equation}
$n_i=b^{\dagger}_ib_i$ is the occupation number operator for {\em holes}\cite{holes} at site $i$ (a hard-core constraint is imposed), whose density is
controlled by the chemical potential $\mu$. Here, only nearest-neighbor (NN) hopping ($t_1$) and interaction ($V_1$) terms are considered
but, as will be discussed later, the effective model derived for the spin model Eq.~\eqref{eq:spin-dimer} further comprises longer-ranged and/or
multi-body couplings.

\begin{figure}
\begin{center}
  \includegraphics*[width=0.33\textwidth,angle=270]{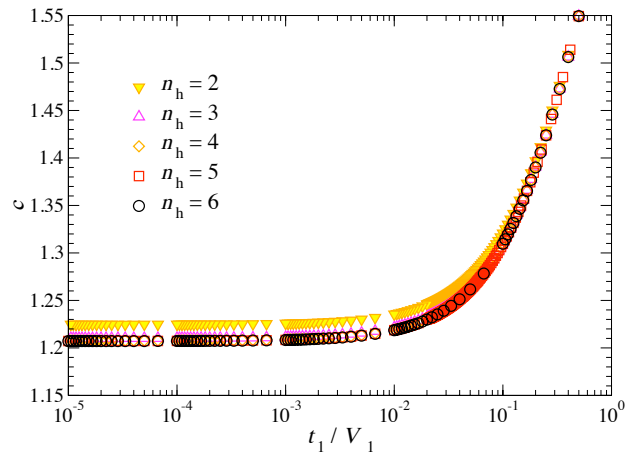}
  \caption{
    (Color online) $c=-E_{\rm DW}/n_{\rm h}t_1$ obtained from EDs
                              on the geometry depicted in Fig.~\ref{fig:DWall}(b), for clusters comprising
                              from $N=8$ ($n_{\rm h}=2$) to $N=24$ ($n_{\rm h}=6$) sites.
  }
  \label{fig:E0_DWall}
  \end{center}
\end{figure}

The model Eq.~\eqref{eq:tVmodel} had been for many years conjectured to support SS phases (see Refs.~\onlinecite{scalettar:95,batrouni:00}
and references therein), but a more systematic numerical analysis\cite{batrouni:00} later disproved earlier evidences in favor of this scenario
and showed that, instead, PS prevents the occurrence of supersolidity. An intuitive explanation for such behavior was put forward
in Ref.~\onlinecite{sengupta:05} by relying on strong coupling arguments. Following this analysis,\cite{sengupta:05} holes [or also particles in the
case of Eq.~\eqref{eq:tVmodel};\cite{holes} however, the effective models to be analyzed later lack particle-hole symmetry] doped into the CBS
ground-state of Eq.~\eqref{eq:tVmodel} for $V_1/t_1 \gg 1$ at half-filling would delocalize with an effective hopping amplitude proportional to
$t_1^{2}/V_1$ and eventually condense, giving origin to SS order. However, this last conclusion is flawed in that it ignores the possibility of PS.
Indeed, in the strongly interacting regime domain wall [DW, depicted as the shaded region in Fig.~\ref{fig:DWall}(a)] formation is energetically
favored for the model Eq.~\eqref{eq:tVmodel}: the energetic gain per {\em doped} hole (we denote the number of {\em doped} holes by $n_{\rm h}$) is
{\em linear} in $t_1$ under these circumstances, $E_{\rm DW}/n_{\rm h} \sim -c t_1$ with $c \in [1,2]$.\cite{sengupta:05}

Since one of our primary goals in the present work is to investigate the interplay between PS and supersolidity, so to be able to decide which
among the two possibilities take place for the spin model herein analyzed [Eq.~\eqref{eq:spin-dimer}], we would like
to obtain a more accurate estimate for $E_{\rm DW}/n_{\rm h}t_1$;  in other words, we would like to pinpoint the actual values assumed by $c \in
[1,2]$. In achieving this goal we completely ignore fluctuations in CBS-ordered regions away from DWs, a supposedly good approximation for
$V_1/t_1 \gg 1$, and consider a simplified $t-V$-like model at half-filling (quarter-filling for {\em doped} holes) defined on the ``comb" geometry
depicted in Fig.~\ref{fig:DWall}(b). In such a simplified model hopping processes with amplitude $t_1$ only take place in between NN sites
linked by the comb's ``teeth" and interaction $V_1$ is only active for holes sitting on NN sites along the ``backbone" [see Fig.~\ref{fig:DWall}(b)].

\begin{figure}
\begin{center}
  \includegraphics*[width=0.125\textwidth,angle=270]{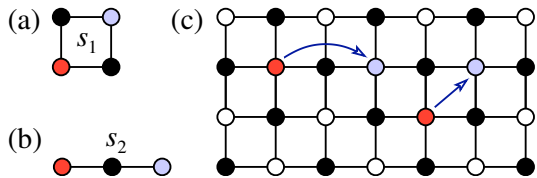}
  \caption{
    (Color online) (a-b) Correlated hoppings behind the ``leapfrog mechanism" for supersolidity
    [holes hop in between red and light-blue sites {\em only} if the dark circles are occupied by holes --- in (a),
    at least one of the sites must be occupied; if both are, the amplitude is $2{s}_{1}$],
    that allow extra holes to delocalize in a CBS background by leapfrogging on the other sublattice (c).
    Adapted from Ref.~\onlinecite{picon:08}.
  }
  \label{fig:leapfrog}
  \end{center}
\end{figure}

In Fig.~\ref{fig:E0_DWall} we plot results for $c=-E_{\rm DW}/n_{\rm h}t_1$, as a function of $t_1/V_1$, obtained from exact diagonalizations (EDs) of the just discussed
simplified model on the comb geometry depicted in Fig.~\ref{fig:DWall}(b), for clusters comprising up to $N=24$ sites (thus, up to $n_{\rm h}=6$ doped holes). We first
notice that $c \approx 1.2$ in the limit of large $V_1/t_1$. On the other hand, since fluctuations away from the DW are ignored in our analysis, we expect our ED results
to {\em underestimate} $c$ for small values of $V_1/t_1$. Nonetheless, we keep this limitation in mind and throughout the remainder of this paper rely on ED results in
estimating the DW energy even in the weakly interacting regime $V_1/t_1 \sim 2$. In doing so, we take advantage of the very small finite size effects in the data shown
in Fig.~\ref{fig:E0_DWall} and obtain $c$ from EDs on finite clusters.

\subsection{Leapfrog Mechanism for Supersolidity}
\label{sec:frog}

In a previous work,\cite{picon:08} we have shown that the instability toward PS is suppressed in models of hard-core bosons on the square lattice that
include, in addition to the terms comprised in the $t-V$ model [Eq.~\eqref{eq:tVmodel}], the correlated hopping processes with amplitudes ${s}_{1}$ and ${s}_{2}$
depicted in Fig.~\ref{fig:leapfrog}. Indeed, such correlated hopping terms have been shown\cite{schmidt:08,picon:08} to favor supersolidity by allowing
doped holes to delocalize on top of a CBS by ``leapfrogging" on the other sub-lattice. We wish now to devise a criterion for determining how large should
the amplitudes ${s}_{1}$ and ${s}_{2}$ be so to inhibit DW formation and thus stabilize a SS. In doing so, we once more rely on strong coupling arguments,
and estimate the energetic gain associated to the leapfrog processes represented in Fig.~\ref{fig:leapfrog} as
\begin{equation}
  E_{\rm SS}/n_{\rm h} = -4\left( 2|{s}_{1}| + |{s}_{2}| \right)~.
  \label{eq:E0_leapfrog}
\end{equation}
That is, $E_{\rm SS}/n_{\rm h}$ is simply the ground-state energy of a single hole doped into a ``frozen" CBS, an approximation expected to hold for
$V_1/t_1 \gg 1$, hopping via the processes with amplitude ${s}_{1}$ and ${s}_{2}$ [Fig.~\ref{fig:leapfrog}(a-b)]. As it happens for our estimate
$E_{\rm DW}/n_{\rm h}=-ct_1$ obtained in Sec.~\ref{sec:DW}, we expect $E_{\rm SS}/n_{\rm h}$ as given by Eq.~\eqref{eq:E0_leapfrog} to underestimate
the actual energetic gain associated to the leapfrog processes.

\begin{figure}
\begin{center}
  \includegraphics*[width=0.15\textwidth,angle=270]{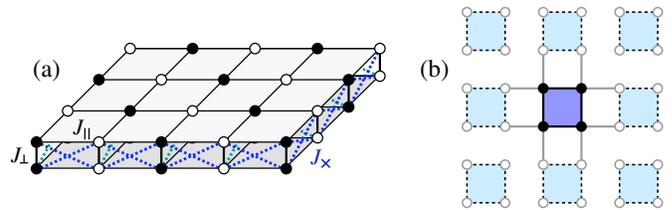}
  \caption{
    (Color online)    (a) Antiferromagnetic bilayer investigated in this paper [Eq.~(\ref{eq:spin-dimer})], with couplings: $J_{\perp}$
    (thick vertical lines), $J_{\parallel}$ (thiner in-layer lines) and $J_{\times}$ (dashed lines). A magnetic field $h$
    promotes singlets (vertical pairs of filled circles) to triplets (pairs of open circles); a CBS configuration
    at half-filling is depicted. (b) $N=2 \times 2$ cluster for SCMFT: interactions (thick black lines) involving only in-cluster sites
    (dark-filled circles) are treated exactly while couplings to the environment (grey lines) in a MF way. Although
    only NN bonds are depicted, the effective model from CORE also includes longer-ranged terms.
  }
  \label{fig:bilayer}
  \end{center}
\end{figure}

We combine the just presented analysis and the one discussed in Sec.~\ref{sec:DW} concerning DW formation and introduce an indicator for analyzing
the interplay between supersolidity and PS: the difference between $E_{\rm SS}/n_{\rm h}$ [Eq.~\eqref{eq:E0_leapfrog}] and $E_{\rm DW}/n_{\rm h}=-ct_1$
(obtained from EDs by using $V_1/t_1$ as input; see Secs.~\ref{sec:DW}), our estimates for the energetic gains respectively
associated to each of these possibilities. Since both estimates are obtained from strong coupling analysis, the indicator $(E_{\rm SS}-E_{\rm DW})/
n_{\rm h}$ can only be expected to be accurate in the limit of $V_1/t_1 \gg 1$. However, we keep this limitation in mind and in the analysis to be
performed in Sec.~\ref{sec:frustrated}, we rely on $(E_{\rm SS}- E_{\rm DW})/n_{\rm h}$ as an indicator even for couplings $V_1/t_1 \sim 2$.

\section{Frustrated Spin Model}
\label{sec:frustrated}

\subsection{Model and Effective Hamiltonian}

In most of the lattice models that have, so far, been shown to display SS properties the effective repulsion (necessary to destabilize the uniform
superfluid and induce a SS state) stems from the presence of nearest-neighbor repulsive terms in XXZ Hamiltonians.\cite{SST,ng:06-laflorencie:07,picon:08}
While these anisotropic models are interesting from a theoretical perspective, their strong anisotropic character renders them unrealistic for antiferromagnetic
Mott insulators. More promising in this sense is the frustrated spin-half Hamiltonian analyzed in Ref.~\onlinecite{chen:10}, defined on a bilayer geometry
[Fig.~\ref{fig:bilayer}(a)]
\begin{equation}
\begin{split}
{\mathcal H} = & \sum_{\left\langle i,j \right\rangle}\left[\sum_{\alpha=1,2} J_{\parallel}\vec{S}_{i,\alpha} \cdot \vec{S}_{j,\alpha}
                          + J_{\times} \left( \vec{S}_{i,1} \cdot \vec{S}_{j,2}+
                          \vec{S}_{i,2} \cdot \vec{S}_{j,1}\right) \right] \\
                          + & \sum_{i} \left[ J_{\perp} \vec{S}_{i,1} \cdot \vec{S}_{i,2} - h \sum_{\alpha=1,2} S^{z}_{i, \alpha} \right]~.
  \label{eq:spin-dimer}
\end{split}
\end{equation}
${\left\langle i,j \right\rangle}$ denotes NN sites in each square layer $\alpha$ of the frustrated
bilayer depicted in Fig.~\ref{fig:bilayer}(a). $J_{\perp}$ couples spins in different layers to
build the basic dimers of the model (we set $J_{\perp}=1$). The applied magnetic
field $h$ acts as a chemical potential, promoting spin-dimers from a singlet (hole) to a triplet (triplon)
state. Effective interactions appear as the result of in-layer $J_{\parallel}$ and frustrating $J_{\times}$
antiferromagnetic couplings. We remark that the lattice depicted in Fig.~\ref{fig:bilayer}(a) remains
invariant if every other spin-dimer is rotated by $\pi$ and thus Eq.~(\ref{eq:spin-dimer}) is
invariant under the transformation $J_{\parallel} \leftrightarrow J_{\times}$, with the consequence
that the phase diagram is symmetric about the line $J_{\parallel}=J_{\times}$.

In studying the model of Eq.~(\ref{eq:spin-dimer}), we adopt an approach similar to the one employed in our previous work Ref.~\onlinecite{picon:08},
where a related unfrustrated model was investigated and to which the reader is referred for details,\cite{holes} and derive an effective bosonic model
by relying on the CORE algorithm.\cite{morningstar:94-96} We consider spin-dimers connected by $J_{\perp}$ as elementary blocks and
select the singlet $\ket{s}=\frac{1}{\sqrt{2}} [\ket{\! \! \uparrow \downarrow \,}-\ket{\! \! \downarrow \uparrow \,}]$ and the $S^{z}=+1$ triplet $\ket{t^{+}} =
\ket{\! \! \uparrow \uparrow \,}$ as the block states in the CORE expansion: for all parameters in Eq.~(\ref{eq:spin-dimer}) considered in the present work,
$J_{\parallel}, J_{\times} \in [0,0.5]$, this choice is justified by the large reduced density-matrix weights associated to such block states and by the rapid
convergence of effective couplings for increasing range in the expansion.\cite{abendschein:07} Effective couplings are derived by diagonalizing clusters
of coupled dimers and by projecting a matching number of low-lying cluster eigenstates onto the basis formed by tensor products of the retained block
states, $\ket{s}$ and $\ket{t^{+}}$.\cite{remark_range2} The effective bosonic Hamiltonian thus obtained is essentially identical to the one derived for the
anisotropic model studied in Ref.~\onlinecite{picon:08}, {\em only} the magnitudes for each coupling being different. Similarly, the effective model obtained
here is {\em not} invariant under particle-hole transformation and, in particular, amplitudes for ``leapfrog processes" are non-zero only when {\em holes} are
involved. From this last observation we expect that only hole-doped SS phases can be stabilized in the spin-dimer model Eq.~(\ref{eq:spin-dimer}) and
conclude that the effective Hamiltonian is more conveniently expressed in terms of hole operators (${n}_{i} = {b}^{\dagger}_{i} {b}_{i}$ is the occupation
number for holes at the dimer-lattice site $\vec{r}_{i}$). We therefore adopt the same notation as in our previous work,\cite{picon:08,holes} to which the
reader is referred for a complete list of single- and multi-hole interactions and hopping processes [see Eqs.~(8, B1-B5) in Ref.~\onlinecite{picon:08}].
We also remark that the effective Hamiltonian preserves the symmetry of the original model Eq.~(\ref{eq:spin-dimer}) and remains invariant under
$J_{\parallel}\leftrightarrow J_{\times}$.

\subsection{Mean-Field Approach}
\label{sec:SCMFT}

Effective Hamiltonians resulting from CORE are often complex and different strategies may be pursued in trying to extract
physically sound results from them. One possibility is the mean-field (MF) theory of Ref.~\onlinecite{picon:08}, which reproduces
semi-quantitatively the results of quantum Monte Carlo (QMC) simulations\cite{ng:06-laflorencie:07} for the anisotropic
spin-dimer model considered therein. However, MF calculations are known to overestimate the extent of SS phases and it would be
desirable to include, at least partially, effects due to quantum fluctuations. From this perspective, the SCMFT\cite{zhao:07} that partially
takes local quantum fluctuations into account and has been recently applied to the $t-V$ model for hard-core bosons on the
triangular lattice\cite{hassan:07} seems particularly well suited for our purposes. Indeed, the extent of the SS phase
in the ground-state phase diagram obtained by applying SCMFT to the $t-V$ model on the triangular lattice\cite{hassan:07}
compares considerably better with results from QMC simulations\cite{SST} than what is found from a more
conventional MF approach.\cite{murthy:97}

\begin{figure}
\begin{center}
  \includegraphics*[width=0.22\textwidth,angle=270]{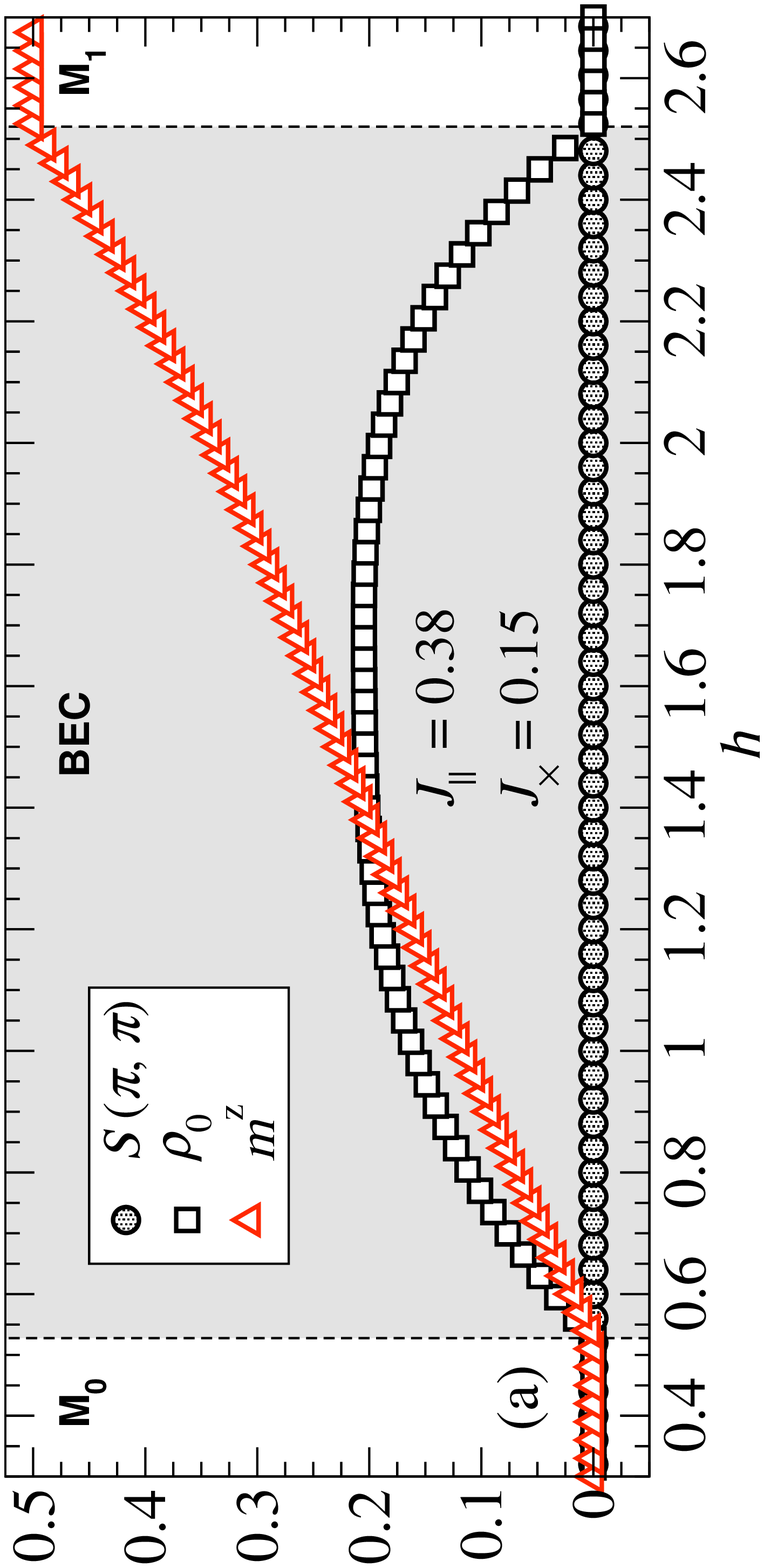}

  \vspace{0.25cm}
  \includegraphics*[width=0.22\textwidth,angle=270]{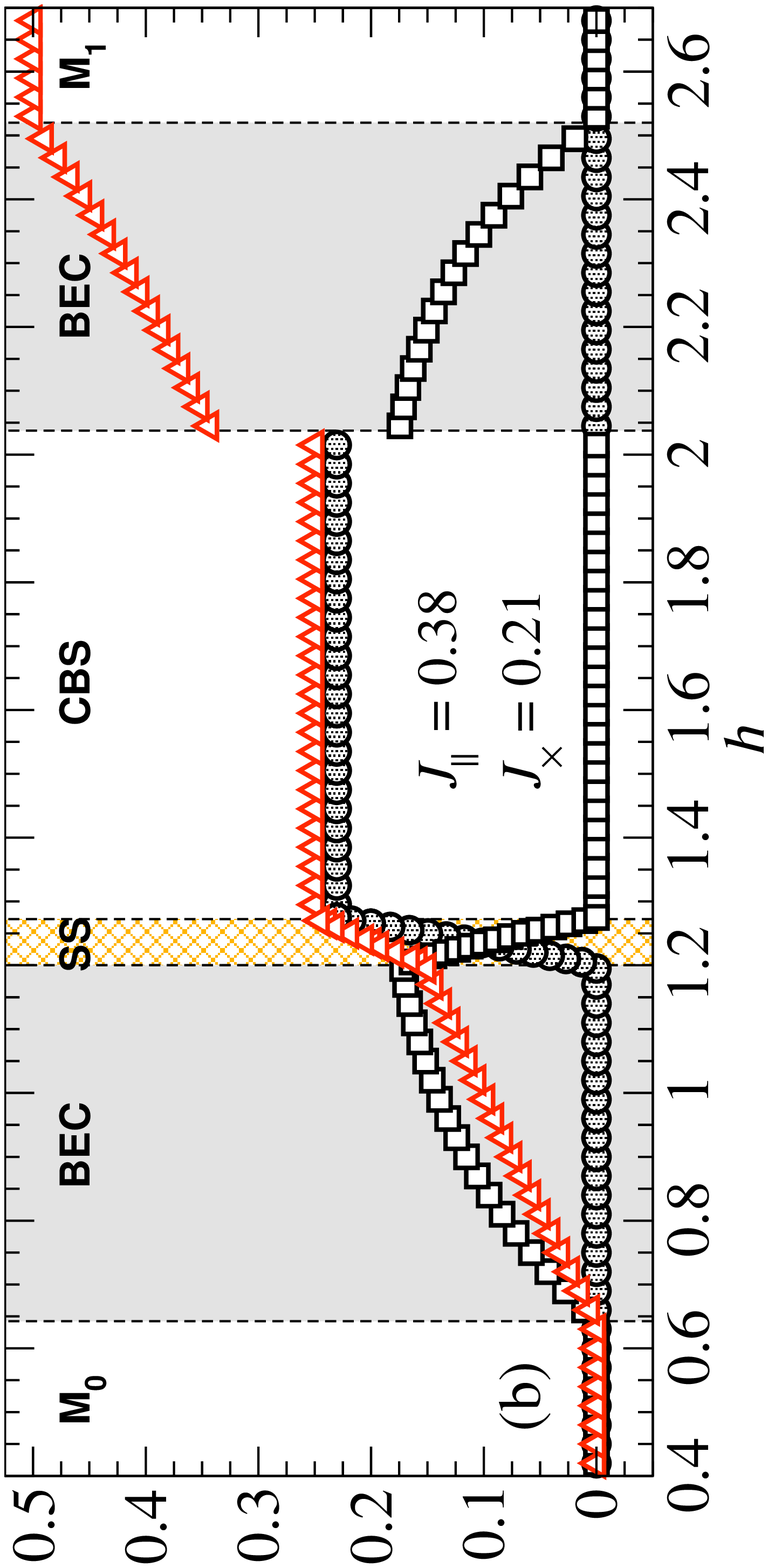}

  \caption{
    (Color online) SCMFT results for the condensate density $\rho_{\rm 0}$ [squares,
    Eq.~(\ref{eq:rho0})], CBS structure factor $S(\pi,\pi)$ [circles, Eq.~(\ref{eq:CBS})]
    and magnetization density [triangles, Eq.~(\ref{eq:mz})]
    for the effective CORE Hamiltonian for Eq.~(\ref{eq:spin-dimer}) with couplings $(J_{\parallel}, J_{\times})$
    considered in Ref.~\onlinecite{chen:10}: (a) $(0.38,0.15)$ and (b) $(0.38,0.21)$. Successive phases for
    increasing magnetic field $h$ are labeled as: spin-gapped (${\rm M}_0$),
    condensate (BEC), supersolid (SS), checkerboard solid (CBS) and fully polarized (${\rm M}_1$).
  }
  \label{fig:MF}
  \end{center}
\end{figure}
\begin{figure}
\begin{center}
  \includegraphics*[width=0.3\textwidth,angle=270]{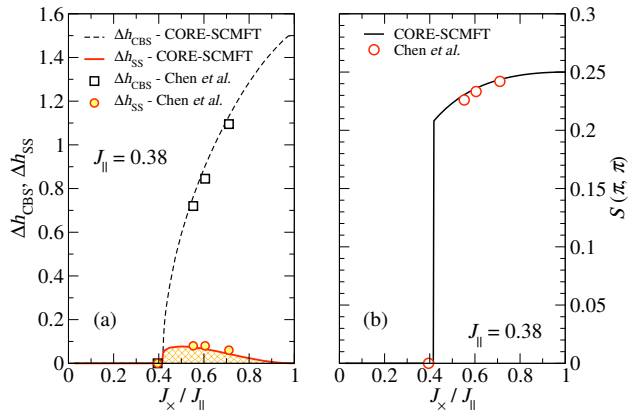}
  \caption{
    (Color online) $J_ {\parallel} =0.38$.
    (a) Extent of CBS ($\Delta h_{\rm CBS}$) and SS ($\Delta h_{\rm SS}$) phases [maximum
    minus minimum value of the field $h$ leading to the corresponding phase for given parameters
    $(J, J_{\times})$]. (b) Value of the structure factor [Eq.~(\ref{eq:CBS})] at the CBS plateau. In (a) and
    (b), symbols indicate results by Chen {\em et al.}\cite{chen:10} and lines the here obtained results.
  }
  \label{fig:extent}
  \end{center}
\end{figure}

SCMFT is applied by diagonalizing the effective CORE Hamiltonian on the $N = 2 \times 2$ cluster depicted
in Fig.~\ref{fig:bilayer}(b). In setting the cluster's Hamiltonian, in-cluster interactions are treated {\em exactly}
while couplings to the environment in a self-consistent way: for instance, a given interaction connecting
sites $\vec{r}_{i}$ and $\vec{r}_{j}$ contributes a term proportional to ${n}_{i} {n}_{j}$ for each in-cluster
bond [thick black lines in Fig.~\ref{fig:bilayer}(b)] and with mean-field terms of the form $[{n}_{i} \langle {n}_{j} \rangle
+ \langle {n}_{i} \rangle {n}_{j}]$ for ``bonds" connecting the cluster to its environment [grey lines in
Fig.~\ref{fig:bilayer}(b)]. At each step, the ground-state for the cluster Hamiltonian is calculated and
expectation values $\langle {n}_{i} \rangle$, $\langle {b}_{i} \rangle$ at every site $\vec{r}_{i}$ computed;
these are then used in setting the mean fields for the next iteration, until convergence is achieved (see Ref.~\onlinecite{hassan:07}
for details). In this way, we compute the condensate density at the point ${\bf k}_0=(\pi,\pi)$
\begin{equation}
\rho_0 = \left| \frac{1}{N} \sum_{j} {\rm e}^{i{\vec{k}}_0\cdot{\vec{r}_{j}}} \langle {b}_{j} \rangle \right|^{2}~,
\label{eq:rho0}
\end{equation}
the CBS structure factor (normalized per site)
\begin{equation}
S(\pi, \pi) = \frac{1}{N^2} \sum_{j,l} {\rm e}^{i{\vec{k}}_0\cdot({\vec{r}}_{j} - {\vec{r}}_{l})}\langle {n}_{j} {n}_{l} \rangle~,
\label{eq:CBS}
\end{equation}
and the  magnetization along the field direction
\begin{equation}
m^{z} = \frac{1}{2N} \sum_{i} \langle ( 1 - {n}_{i} ) \rangle~.
\label{eq:mz}
\end{equation}

In Fig.~\ref{fig:MF} we plot these quantities
as a function of the magnetic field $h$ for couplings considered in Ref.~\onlinecite{chen:10}, $(J_{\parallel}, J_{\times})=(0.38,0.15)$
and $(0.38,0.21)$. We first notice that the overall agreement between our results and the data presented in
Ref.~\onlinecite{chen:10} is remarkably good.\cite{agreement} For the least frustrated case of $(J_{\parallel}, J_{\times})=(0.38,0.15)$
[Fig.~\ref{fig:MF}(a)], the system first undergoes a quantum transition from a spin-gapped (equivalent to a trivial bosonic Mott
insulator with zero-filling for triplons, ${\rm M_0}$ in Fig.~\ref{fig:MF}) to a BEC phase at the lower critical field $h_{\rm c1}$,
and then from the BEC to a fully polarized phase (trivial Mott insulator with unitary triplon filling, ${\rm M_1}$) at the upper critical
field $h_{\rm c2}$. More interestingly, additional CBS and SS phases are stabilized for the more frustrated case of
$(J_{\parallel}, J_{\times})=(0.38,0.21)$ [Fig.~\ref{fig:MF}(b)]. The existence of a SS phase at the low-field boundary of the
CBS plateau, with finite values for both $\rho_0$ and $S(\pi, \pi)$ is at least partially due to the presence of correlated hoppings
for {\em holes} in the effective CORE Hamiltonian. Indeed, no SS phase is observed for an ``effective model" obtained by
setting ${s}_{1}= {s}_{2}=0$ [Fig.~\ref{fig:leapfrog}(a-b)] while keeping all the other effective couplings unchanged.
This situation is to be contrasted with the first-order transition from CBS to BEC at higher fields, explained by the {\em vanishing}
amplitudes for correlated hoppings for {\em triplons} for all values $J_{\parallel}, J_{\times} \in [0,0.5]$.

We proceed by varying the frustrating coupling $J_{\times}$ while fixing $J_{\parallel}=0.38$. In Fig.~\ref{fig:extent}(a)
we plot the extent of the SS, CBS phase, respectively $\Delta h_{\rm SS, CBS} = h_{\rm SS, CBS}^{max} - h_{\rm SS, CBS}^{min}$
[$h_ {\rm SS, CBS}^{max}$ ($h_ {\rm SS, CBS}^{min}$) denotes the upper (lower) boundary of the SS, CBS phase] as a function of
$J_{\times}/J_{\parallel}$. In order to further gauge the accuracy of the here employed CORE-SCMFT approach in Fig.~\ref{fig:extent}(a)
our results for $\Delta h_{\rm SS}$ and $\Delta h_{\rm CBS}$ are compared against those from Ref.~\onlinecite{chen:10} and in Fig.~\ref{fig:extent}(b)
we plot both our results and those from Ref.~\onlinecite{chen:10} for the structure factor $S(\pi,\pi)$ at the CBS plateau. Excellent agreement is
found in both cases and we further remark that our results for $S(\pi,\pi)$ in the CBS phase in Fig.~\ref{fig:extent}(b) confirm that quantum
fluctuations are indeed partially taken into account by SCMFT: in contrast to what happens with the semi-classical MF approach employed
in Ref.~\onlinecite{picon:08}, here the value of $S(\pi,\pi)$ at the plateau is somewhat reduced from its classical value $S_{\rm classical}(\pi,\pi)=1/4$.

At this point, and despite of its aforementioned attractive features, it is important to have in mind an important limitation of the here employed SCMFT
procedure: since calculations rely on diagonalizations of a $2 \times 2$ cluster [Fig.~\ref{fig:bilayer}(b)], only {\em homogeneous} solutions, displaying
order consistent with at most quadrupling of the unit cell, are obtainable. This excludes {\em inhomogeneous} solutions such as those associated
with the presence of DWs [Fig.~\ref{fig:DWall}(a)] and has the consequence that our combined CORE-SCMFT approach is insensitive to the occurrence
of PS. In what follows, we rely on the strong coupling analysis presented in Sec.~\ref{sec:PS} and analyze the interplay between PS and spin-SS order
in the phase diagram of the model Eq.~(\ref{eq:spin-dimer}).

\subsection{Phase Diagram and Phase Separation}
\label{sec:phase_diag}

\begin{figure}
\begin{center}
  \includegraphics*[width=0.45\textwidth]{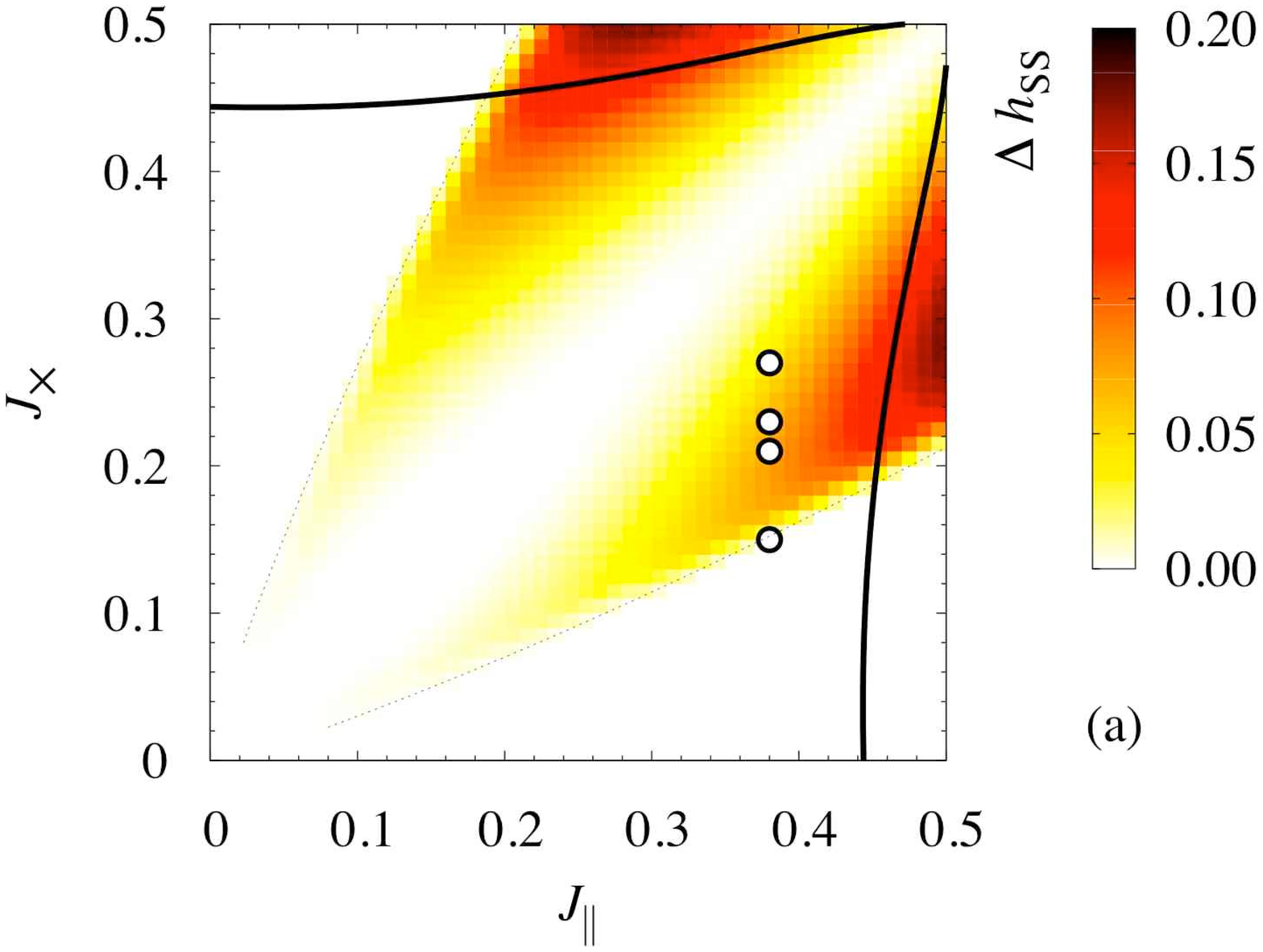}
   \vspace{0.25cm}
  \includegraphics*[width=0.45\textwidth]{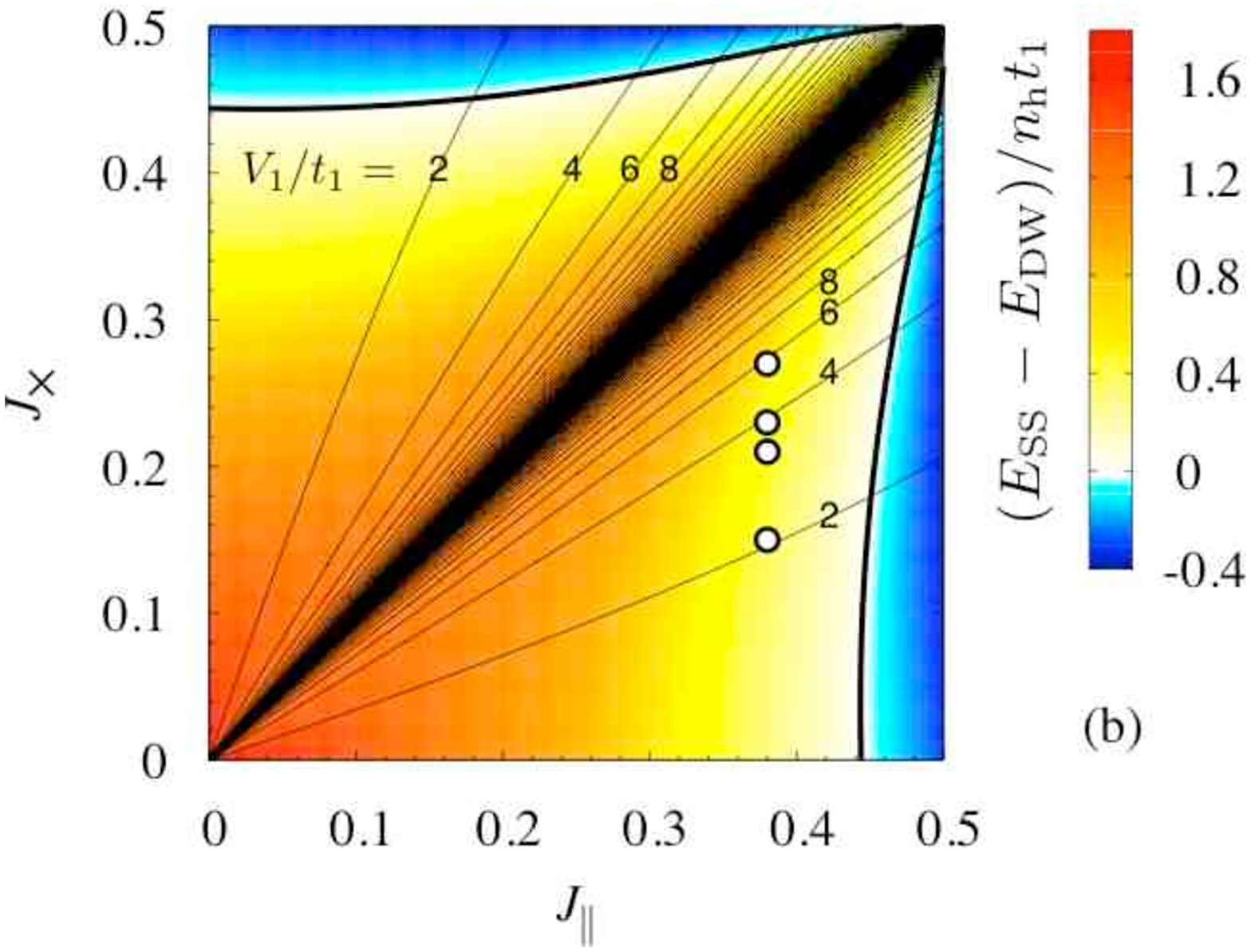} 

  \caption{(Color online) (a) SCMFT results for the extent of the SS phase $\Delta h_{\rm SS}$ (see main text) for the frustrated DAF Eq.~(\ref{eq:spin-dimer}).
                  The symmetry $J_{\parallel} \leftrightarrow J_{\times}$ has been explored in obtaining the data. Regions where supersolidity [PS] is expected, where
                  $(E_{\rm SS}-E_{\rm DW})/n_{\rm h}t_1<0$ [$(E_{\rm SS}-E_{\rm DW})/n_{\rm h}t_1>0$] are marked by the label SS [PS]. Dashed lines indicate
                  threshold values for a CBS/SS to appear at the mean-field level. (b) $(E_{\rm SS}-E_{\rm DW})
                  /n_{\rm h}t_1$, as obtained from EDs on an $N=16$ ($n_{\rm h}=4$ doped holes) site cluster with the comb geometry depicted in Fig.~\ref{fig:DWall}(b),
                  for the model Eq.~(\ref{eq:spin-dimer}). Contour levels for $V_1/t_1$ (obtained from the CORE expansion) are indicated by thin lines and the values
                  $V_1/t_1=2$, $4$, $6$ and $8$ are highlighted. In both panels, circles indicate couplings investigated by Chen {\it{et al.}}\cite{chen:10} and the thick
                  line couplings yielding the threshold value $(E_{\rm SS}-E_{\rm DW})/n_{\rm h}t_1=0$.
                  }
  \label{fig:range}
  \end{center}
\end{figure}

We now turn our attention to the obtention of a global $J_{\parallel}$ --- $J_{\times}$ phase diagram that may guide the experimental search for realizations of
spin-supersolidity and therefore extend our analysis by varying $J_{\parallel}$ in Eq.~(\ref{eq:spin-dimer}). In Fig.~\ref{fig:range}(a) we plot $\Delta h_{\rm SS}$
as a function of $J_{\parallel}, J_{\times} \in [0,0.5]$, obtained from the combined CORE-SCMFT procedure. These results suggest that, far from being a rare
occurrence, spin-supersolidity is widespread throughout the parameter space and can extend over fairly wider ranges of $h$ than it is observed for the value
$J_{\parallel}=0.38$ [couplings considered in Ref.~\onlinecite{chen:10} are highlighted in Fig.~\ref{fig:range}(a)]. However, under the light of our discussion
in Sec.~\ref{sec:PS} concerning PS in systems of hard-core bosons on a lattice, some caution is required in drawing conclusions from the results shown in
Fig.~\ref{fig:range}(a).

Following the discussion in Sec.~\ref{sec:PS}, we evaluate $(E_{\rm SS}-E_{\rm DW})/n_{\rm h}t_1$, our indicator for analyzing the interplay between
PS and supersolidity for hard-core bosons on the square lattice, as a function of $J_{\parallel}, J_{\times} \in [0,0.5]$ for Eq.~(\ref{eq:spin-dimer}).
$E_{\rm SS}/n_{\rm h}$, our strong coupling estimate for the energetic gain associated to occurrence of SS order, is readily obtained by plugging
the amplitudes for the leapfrog processes ${s}_{1}$ and ${s}_{2}$ [Fig.~\ref{fig:leapfrog}(a-b)] obtained from CORE into Eq.~\eqref{eq:E0_leapfrog}.
On the other hand, the estimate $E_{\rm DW}/n_{\rm h}$ for the energy associated to PS is obtained from numerical EDs for the simplified model for DWs
defined on the ``comb geometry" discussed in Sec.~\ref{sec:DW} by using the effective ratio $V_1/t_1$, as obtained from the CORE expansion for each set
$J_{\parallel}, J_{\times}$ in Eq.~(\ref{eq:spin-dimer}), as an input. EDs are performed on a small cluster comprising $N=16$ sites ($n_{\rm h}=4$
doped holes): as mentioned in Sec.~\ref{sec:DW}, this is justified by the absence of sizable finite size effects for the data displayed in Fig.~\ref{fig:E0_DWall}.

We plot $(E_{\rm SS}-E_{\rm DW})/n_{\rm h}t_1$ as a function of $J_{\parallel}, J_{\times} \in [0,0.5]$ in Fig.~\ref{fig:range}(b) and assume that two
conditions must be simultaneously fulfilled for SS phases to exist for the model Eq.~(\ref{eq:spin-dimer}): (i) a SS must be observed within CORE-SCMFT
{\em and} (ii) $(E_{\rm SS}-E_{\rm DW})/n_{\rm h}t_1<0$. Values of $J_{\parallel}, J_{\times}$ leading to $(E_{\rm SS}-E_{\rm DW})/n_{\rm h}t_1=0$,
the threshold value for a SS phase to appear, are indicated by the thick continuous curve in Fig.~\ref{fig:range}. We notice that not all values of $J_{\parallel},
J_{\times}$ yielding a spin-SS phase within our CORE-SCMFT approach fulfill $(E_{\rm SS}-E_{\rm DW})/n_{\rm h}t_1<0$ and expect PS to take place under
these circumstances instead. Despite of the fact that the condition $(E_{\rm SS}-E_{\rm DW})/n_{\rm h}t_1<0$ considerably shrinks the size of the region expected
to support SS phases from a pure CORE-SCMFT analysis, supersolidity is still observed for a wide range of couplings in Eq.~(\ref{eq:spin-dimer}) [Fig.~\ref{fig:range}(a)],
possibly realizable in real magnets.

Intriguingly, we notice that the parameters (circles in Fig.~\ref{fig:range}) for which a spin-SS phase has been detected by Chen {\em et al}.,\cite{chen:10} and also by
the pure CORE-SCMFT analysis devised here (Figs.~\ref{fig:MF} and \ref{fig:extent}), fail to satisfy $(E_{\rm SS}-E_{\rm DW})/n_{\rm h}t_1<0$ [Fig.~\ref{fig:range}(b)].
Although we cannot exclude the possibility that our criterion, that rigorously applies only in the limit $V_1/t_1 \gg 1$, is too stringent for the frustrated model
Eq.~(\ref{eq:spin-dimer}), we remark that a SS phase is obtained within our CORE-SCMFT approach\cite{1st_order} and in Ref.~\onlinecite{chen:10} even for
couplings $J_{\parallel} \sim J_{\times}$, where our strong coupling arguments become accurate [the ratio ${V}_{1}/t_{1}$ diverges toward the line $J_{\parallel} =
J_{\times}$; see Fig.~\ref{fig:range}(b), where contour levels for ${V}_{1}/{t}_{1}$ are plotted as thin continuous lines]. This inconsistency suggests that both
CORE-SCMFT and the novel algorithm employed in Ref.~\onlinecite{chen:10} are insensitive to the instability toward PS in systems of hard-core bosons on the
square lattice and that the obtention of SS phase for $(E_{\rm SS}-E_{\rm DW})/n_{\rm h}t_1>0$ is {\em spurious}.\cite{1st_order} It would therefore be important
to further test the ability of the algorithm employed in Ref.~\onlinecite{chen:10} to detect PS in bosonic lattice models by, for instance, checking how it compares to
QMC for the unfrustrated model studied in Refs.~\onlinecite{ng:06-laflorencie:07,picon:08} regarding this issue.

\section{Summary}
\label{sec:summary}

Summarizing, we have studied a spin-half frustrated bilayer model by combining CORE and SCMFT. Our results reveal the presence of a spin-SS phase under
applied magnetic field, which appears at the edge of a half-saturated magnetization plateau and is stabilized by a ``leapfrog mechanism".\cite{picon:08} We address
the interplay between supersolidity and instability toward PS, that precludes the emergence of spin-SS phases, by devising a quantitative criterion based on strong
coupling arguments. This criterion is generically applicable to systems of hard-core bosons on the square lattice, and it would be interesting to further assess its
validity by investigating models where the interplay between PS and SS can be independently analyzed. By relying on this criterion, we obtain a global phase
diagram for the frustrated spin-dimer antiferromagnet considered herein, and show that a spin-SS phase is stable against PS for couplings realizable in real
magnets. We expect that our results may guide the experimental search for systems exhibiting spin-supersolidity.

\begin{acknowledgments}
We acknowledge fruitful discussions with C.~D.~Batista and M. Troyer,
as well as funding from the French ANR program ANR-08-JCJC-0056-01,
from ARC (Australia), from the SNF and from MaNEP (Switzerland). NL acknowledges LPT Toulouse for hospitality.

\end{acknowledgments}


\begin{thebibliography}{41}
\expandafter\ifx\csname natexlab\endcsname\relax\def\natexlab#1{#1}\fi
\expandafter\ifx\csname bibnamefont\endcsname\relax
  \def\bibnamefont#1{#1}\fi
\expandafter\ifx\csname bibfnamefont\endcsname\relax
  \def\bibfnamefont#1{#1}\fi
\expandafter\ifx\csname citenamefont\endcsname\relax
  \def\citenamefont#1{#1}\fi
\expandafter\ifx\csname url\endcsname\relax
  \def\url#1{\texttt{#1}}\fi
\expandafter\ifx\csname urlprefix\endcsname\relax\def\urlprefix{URL }\fi
\providecommand{\bibinfo}[2]{#2}
\providecommand{\eprint}[2][]{\url{#2}}

\bibitem[{\citenamefont{Giamarchi et~al.}(2008)\citenamefont{Giamarchi,
  R\"uegg, and Tchernyshyov}}]{giamarchi:08}
\bibinfo{author}{\bibfnamefont{T.}~\bibnamefont{Giamarchi}},
  \bibinfo{author}{\bibfnamefont{C.}~\bibnamefont{R\"uegg}}, \bibnamefont{and}
  \bibinfo{author}{\bibfnamefont{O.}~\bibnamefont{Tchernyshyov}},
  \bibinfo{journal}{Nat. Phys.} \textbf{\bibinfo{volume}{4}},
  \bibinfo{pages}{198} (\bibinfo{year}{2008}).

\bibitem[{\citenamefont{Mila}(1998)}]{Mila98}
\bibinfo{author}{\bibfnamefont{F.}~\bibnamefont{Mila}}, \bibinfo{journal}{Eur.
  Phys. J. B} \textbf{\bibinfo{volume}{6}}, \bibinfo{pages}{201}
  (\bibinfo{year}{1998}).

\bibitem[{\citenamefont{Totsuka}(1998)}]{Totsuka98-Giamarchi99}
\bibinfo{author}{\bibfnamefont{K.}~\bibnamefont{Totsuka}},
  \bibinfo{journal}{Phys. Rev. B} \textbf{\bibinfo{volume}{57}},
  \bibinfo{pages}{3454} (\bibinfo{year}{1998});
%
\bibinfo{author}{\bibfnamefont{T.}~\bibnamefont{Giamarchi}} \bibnamefont{and}
  \bibinfo{author}{\bibfnamefont{A.~M.} \bibnamefont{Tsvelik}},
  \bibinfo{journal}{Phys. Rev. B} \textbf{\bibinfo{volume}{59}},
  \bibinfo{pages}{11398} (\bibinfo{year}{1999}).

\bibitem[{\citenamefont{Sebastian et~al.}(2006)\citenamefont{Sebastian,
  Harrison, Batista, Balicas, Jaime, Sharma, Kawashima, and
  Fisher}}]{sebastian:06}
\bibinfo{author}{\bibfnamefont{S.~E.} \bibnamefont{Sebastian}},
  \bibinfo{author}{\bibfnamefont{N.}~\bibnamefont{Harrison}},
  \bibinfo{author}{\bibfnamefont{C.~D.} \bibnamefont{Batista}},
  \bibinfo{author}{\bibfnamefont{L.}~\bibnamefont{Balicas}},
  \bibinfo{author}{\bibfnamefont{M.}~\bibnamefont{Jaime}},
  \bibinfo{author}{\bibfnamefont{P.~A.} \bibnamefont{Sharma}},
  \bibinfo{author}{\bibfnamefont{N.}~\bibnamefont{Kawashima}},
  \bibnamefont{and} \bibinfo{author}{\bibfnamefont{I.~R.}
  \bibnamefont{Fisher}}, \bibinfo{journal}{Nature}
  \textbf{\bibinfo{volume}{441}}, \bibinfo{pages}{617} (\bibinfo{year}{2006}).

\bibitem[{\citenamefont{R\"uegg et~al.}(2007)\citenamefont{R\"uegg, McMorrow,
  Normand, R\o{}nnow, Sebastian, Fisher, Batista, Gvasaliya, Niedermayer, and
  Stahn}}]{Ruegg07}
\bibinfo{author}{\bibfnamefont{C.}~\bibnamefont{R\"uegg}},
  \bibinfo{author}{\bibfnamefont{D.~F.} \bibnamefont{McMorrow}},
  \bibinfo{author}{\bibfnamefont{B.}~\bibnamefont{Normand}},
  \bibinfo{author}{\bibfnamefont{H.~M.} \bibnamefont{R\o{}nnow}},
  \bibinfo{author}{\bibfnamefont{S.~E.} \bibnamefont{Sebastian}},
  \bibinfo{author}{\bibfnamefont{I.~R.} \bibnamefont{Fisher}},
  \bibinfo{author}{\bibfnamefont{C.~D.} \bibnamefont{Batista}},
  \bibinfo{author}{\bibfnamefont{S.~N.} \bibnamefont{Gvasaliya}},
  \bibinfo{author}{\bibfnamefont{C.}~\bibnamefont{Niedermayer}},
  \bibnamefont{and} \bibinfo{author}{\bibfnamefont{J.}~\bibnamefont{Stahn}},
  \bibinfo{journal}{Phys. Rev. Lett.} \textbf{\bibinfo{volume}{98}},
  \bibinfo{pages}{017202} (\bibinfo{year}{2007}).

\bibitem[{\citenamefont{{Kr{\"a}mer} et~al.}(2007)\citenamefont{{Kr{\"a}mer},
  {Stern}, {Horvati{\'c}}, {Berthier}, {Kimura}, and {Fisher}}}]{Kramer07}
\bibinfo{author}{\bibfnamefont{S.}~\bibnamefont{{Kr{\"a}mer}}},
  \bibinfo{author}{\bibfnamefont{R.}~\bibnamefont{{Stern}}},
  \bibinfo{author}{\bibfnamefont{M.}~\bibnamefont{{Horvati{\'c}}}},
  \bibinfo{author}{\bibfnamefont{C.}~\bibnamefont{{Berthier}}},
  \bibinfo{author}{\bibfnamefont{T.}~\bibnamefont{{Kimura}}}, \bibnamefont{and}
  \bibinfo{author}{\bibfnamefont{I.~R.} \bibnamefont{{Fisher}}},
  \bibinfo{journal}{Phys. Rev. B} \textbf{\bibinfo{volume}{76}},
  \bibinfo{pages}{100406} (\bibinfo{year}{2007}).

\bibitem[{\citenamefont{Laflorencie and Mila}(2009)}]{Laflorencie09}
\bibinfo{author}{\bibfnamefont{N.}~\bibnamefont{Laflorencie}} \bibnamefont{and}
  \bibinfo{author}{\bibfnamefont{F.}~\bibnamefont{Mila}},
  \bibinfo{journal}{Phys. Rev. Lett.} \textbf{\bibinfo{volume}{102}},
  \bibinfo{pages}{060602} (\bibinfo{year}{2009}).

\bibitem[{\citenamefont{Takigawa and Mila}(2010)}]{Takigawa-Mila10}
\bibinfo{author}{\bibfnamefont{M.}~\bibnamefont{Takigawa}} \bibnamefont{and}
  \bibinfo{author}{\bibfnamefont{F.}~\bibnamefont{Mila}}, in
  \emph{\bibinfo{booktitle}{{Introduction to Frustrated Magnetism}}}, edited by
  \bibinfo{editor}{\bibfnamefont{C.}~\bibnamefont{{Lacroix}}},
  \bibinfo{editor}{\bibfnamefont{P.}~\bibnamefont{{Mendels}}},
  \bibnamefont{and} \bibinfo{editor}{\bibfnamefont{F.}~\bibnamefont{{Mila}}}
  (\bibinfo{publisher}{Springer - Berlin}, \bibinfo{year}{2011}).

\bibitem[{\citenamefont{Kageyama et~al.}(1999)\citenamefont{Kageyama,
  Yoshimura, Stern, Mushnikov, Onizuka, Kato, Kosuge, Slichter, Goto, and
  Ueda}}]{kageyama:99}
\bibinfo{author}{\bibfnamefont{H.}~\bibnamefont{Kageyama}},
  \bibinfo{author}{\bibfnamefont{K.}~\bibnamefont{Yoshimura}},
  \bibinfo{author}{\bibfnamefont{R.}~\bibnamefont{Stern}},
  \bibinfo{author}{\bibfnamefont{N.~V.} \bibnamefont{Mushnikov}},
  \bibinfo{author}{\bibfnamefont{K.}~\bibnamefont{Onizuka}},
  \bibinfo{author}{\bibfnamefont{M.}~\bibnamefont{Kato}},
  \bibinfo{author}{\bibfnamefont{K.}~\bibnamefont{Kosuge}},
  \bibinfo{author}{\bibfnamefont{C.~P.} \bibnamefont{Slichter}},
  \bibinfo{author}{\bibfnamefont{T.}~\bibnamefont{Goto}}, \bibnamefont{and}
  \bibinfo{author}{\bibfnamefont{Y.}~\bibnamefont{Ueda}},
  \bibinfo{journal}{Phys. Rev. Lett.} \textbf{\bibinfo{volume}{82}},
  \bibinfo{pages}{3168} (\bibinfo{year}{1999}).

\bibitem[{\citenamefont{{Kodama} et~al.}(2002)\citenamefont{{Kodama},
  {Takigawa}, {Horvati\'c}, {Berthier}, {Kageyama}, {Ueda}, {Miyahara},
  {Becca}, and {Mila}}}]{kodama:02}
\bibinfo{author}{\bibfnamefont{K.}~\bibnamefont{{Kodama}}},
  \bibinfo{author}{\bibfnamefont{M.}~\bibnamefont{{Takigawa}}},
  \bibinfo{author}{\bibfnamefont{M.}~\bibnamefont{{Horvati\'c}}},
  \bibinfo{author}{\bibfnamefont{C.}~\bibnamefont{{Berthier}}},
  \bibinfo{author}{\bibfnamefont{H.}~\bibnamefont{{Kageyama}}},
  \bibinfo{author}{\bibfnamefont{Y.}~\bibnamefont{{Ueda}}},
  \bibinfo{author}{\bibfnamefont{S.}~\bibnamefont{{Miyahara}}},
  \bibinfo{author}{\bibfnamefont{F.}~\bibnamefont{{Becca}}}, \bibnamefont{and}
  \bibinfo{author}{\bibfnamefont{F.}~\bibnamefont{{Mila}}},
  \bibinfo{journal}{Science} \textbf{\bibinfo{volume}{298}},
  \bibinfo{pages}{395} (\bibinfo{year}{2002}).

\bibitem[{\citenamefont{Dorier et~al.}(2008)\citenamefont{Dorier, Schmidt, and
  Mila}}]{Dorier08-Abendschein08}
\bibinfo{author}{\bibfnamefont{J.}~\bibnamefont{Dorier}},
  \bibinfo{author}{\bibfnamefont{K.~P.} \bibnamefont{Schmidt}},
  \bibnamefont{and} \bibinfo{author}{\bibfnamefont{F.}~\bibnamefont{Mila}},
  \bibinfo{journal}{Phys. Rev. Lett.} \textbf{\bibinfo{volume}{101}},
  \bibinfo{pages}{250402} (\bibinfo{year}{2008});
%
\bibinfo{author}{\bibfnamefont{A.}~\bibnamefont{{Abendschein}}}
  \bibnamefont{and}
  \bibinfo{author}{\bibfnamefont{S.}~\bibnamefont{{Capponi}}},
  \bibinfo{journal}{Phys. Rev. Lett.} \textbf{\bibinfo{volume}{101}},
  \bibinfo{pages}{227201} (\bibinfo{year}{2008}).

\bibitem[{\citenamefont{{Andreev} and {Lifshitz}}(1969)}]{andreev:69}
\bibinfo{author}{\bibfnamefont{A.~F.} \bibnamefont{{Andreev}}}
  \bibnamefont{and} \bibinfo{author}{\bibfnamefont{I.~M.}
  \bibnamefont{{Lifshitz}}}, \bibinfo{journal}{Sov. Phys. JETP}
  \textbf{\bibinfo{volume}{29}}, \bibinfo{pages}{1107} (\bibinfo{year}{1969}).

\bibitem[{\citenamefont{{Kim} and {Chan}}(2004{\natexlab{a}})}]{kim:04}
\bibinfo{author}{\bibfnamefont{E.}~\bibnamefont{{Kim}}} \bibnamefont{and}
  \bibinfo{author}{\bibfnamefont{M.~H.~W.} \bibnamefont{{Chan}}},
  \bibinfo{journal}{Nature} \textbf{\bibinfo{volume}{427}},
  \bibinfo{pages}{225} (\bibinfo{year}{2004}{\natexlab{a}});
%
\bibinfo{author}{\bibfnamefont{E.}~\bibnamefont{{Kim}}} \bibnamefont{and}
  \bibinfo{author}{\bibfnamefont{M.~H.~W.} \bibnamefont{{Chan}}},
  \bibinfo{journal}{Science} \textbf{\bibinfo{volume}{305}},
  \bibinfo{pages}{1941} (\bibinfo{year}{2004}{\natexlab{b}}).

\bibitem[{\citenamefont{Balibar}(2010)}]{Balibar10}
\bibinfo{author}{\bibfnamefont{S.}~\bibnamefont{Balibar}},
  \bibinfo{journal}{Nature} \textbf{\bibinfo{volume}{464}},
  \bibinfo{pages}{176} (\bibinfo{year}{2010}).

\bibitem[{\citenamefont{{Senthil} et~al.}(2004)\citenamefont{{Senthil},
  {Vishwanath}, {Balents}, {Sachdev}, and {Fisher}}}]{senthil:04}
\bibinfo{author}{\bibfnamefont{T.}~\bibnamefont{{Senthil}}},
  \bibinfo{author}{\bibfnamefont{A.}~\bibnamefont{{Vishwanath}}},
  \bibinfo{author}{\bibfnamefont{L.}~\bibnamefont{{Balents}}},
  \bibinfo{author}{\bibfnamefont{S.}~\bibnamefont{{Sachdev}}},
  \bibnamefont{and} \bibinfo{author}{\bibfnamefont{M.~P.~A.}
  \bibnamefont{{Fisher}}}, \bibinfo{journal}{Science}
  \textbf{\bibinfo{volume}{303}}, \bibinfo{pages}{1490} (\bibinfo{year}{2004}).

\bibitem[{\citenamefont{{Momoi} and {Totsuka}}(2000)}]{momoi:00}
\bibinfo{author}{\bibfnamefont{T.}~\bibnamefont{{Momoi}}} \bibnamefont{and}
  \bibinfo{author}{\bibfnamefont{K.}~\bibnamefont{{Totsuka}}},
  \bibinfo{journal}{Phys. Rev. B} \textbf{\bibinfo{volume}{62}},
  \bibinfo{pages}{15067} (\bibinfo{year}{2000}).

\bibitem[{\citenamefont{Ng and Lee}(2006)}]{ng:06-laflorencie:07}
\bibinfo{author}{\bibfnamefont{K.-K.} \bibnamefont{Ng}} \bibnamefont{and}
  \bibinfo{author}{\bibfnamefont{T.~K.} \bibnamefont{Lee}},
  \bibinfo{journal}{Phys. Rev. Lett.} \textbf{\bibinfo{volume}{97}},
  \bibinfo{pages}{127204} (\bibinfo{year}{2006});
%
\bibinfo{author}{\bibfnamefont{N.}~\bibnamefont{Laflorencie}} \bibnamefont{and}
  \bibinfo{author}{\bibfnamefont{F.}~\bibnamefont{Mila}},
  \bibinfo{journal}{Phys. Rev. Lett} \textbf{\bibinfo{volume}{99}},
  \bibinfo{pages}{027202} (\bibinfo{year}{2007}).

\bibitem[{\citenamefont{Picon et~al.}(2008)\citenamefont{Picon, Albuquerque,
  Schmidt, Laflorencie, Troyer, and Mila}}]{picon:08}
\bibinfo{author}{\bibfnamefont{J.-D.} \bibnamefont{Picon}},
  \bibinfo{author}{\bibfnamefont{A.~F.} \bibnamefont{Albuquerque}},
  \bibinfo{author}{\bibfnamefont{K.~P.} \bibnamefont{Schmidt}},
  \bibinfo{author}{\bibfnamefont{N.}~\bibnamefont{Laflorencie}},
  \bibinfo{author}{\bibfnamefont{M.}~\bibnamefont{Troyer}}, \bibnamefont{and}
  \bibinfo{author}{\bibfnamefont{F.}~\bibnamefont{Mila}},
  \bibinfo{journal}{Phys. Rev. B} \textbf{\bibinfo{volume}{78}},
  \bibinfo{pages}{184418} (\bibinfo{year}{2008}).

\bibitem[{\citenamefont{Sengupta and Batista}(2007)}]{sengupta:07a}
\bibinfo{author}{\bibfnamefont{P.}~\bibnamefont{Sengupta}} \bibnamefont{and}
  \bibinfo{author}{\bibfnamefont{C.~D.} \bibnamefont{Batista}},
  \bibinfo{journal}{Phys. Rev. Lett.} \textbf{\bibinfo{volume}{98}},
  \bibinfo{pages}{227201} (\bibinfo{year}{2007}).

\bibitem[{\citenamefont{Chen et~al.}(2010)\citenamefont{Chen, Lai, and
  Yang}}]{chen:10}
\bibinfo{author}{\bibfnamefont{P.}~\bibnamefont{Chen}},
  \bibinfo{author}{\bibfnamefont{C.-Y.} \bibnamefont{Lai}}, \bibnamefont{and}
  \bibinfo{author}{\bibfnamefont{M.-F.} \bibnamefont{Yang}},
  \bibinfo{journal}{Phys. Rev. B} \textbf{\bibinfo{volume}{81}},
  \bibinfo{pages}{020409} (\bibinfo{year}{2010}).

\bibitem[{\citenamefont{{Morningstar} and {Weinstein}}(1994)}]{morningstar:94-96}
\bibinfo{author}{\bibfnamefont{C.~J.} \bibnamefont{{Morningstar}}}
  \bibnamefont{and}
  \bibinfo{author}{\bibfnamefont{M.}~\bibnamefont{{Weinstein}}},
  \bibinfo{journal}{Phys. Rev. Lett.} \textbf{\bibinfo{volume}{73}},
  \bibinfo{pages}{1873} (\bibinfo{year}{1994});
%
\bibinfo{author}{\bibfnamefont{C.~J.} \bibnamefont{{Morningstar}}}
  \bibnamefont{and}
  \bibinfo{author}{\bibfnamefont{M.}~\bibnamefont{{Weinstein}}},
  \bibinfo{journal}{Phys. Rev. D} \textbf{\bibinfo{volume}{54}},
  \bibinfo{pages}{4131} (\bibinfo{year}{1996}).

\bibitem[{hol()}]{holes}
\bibinfo{note}{The model Eq.~\eqref{eq:tVmodel} displays particle-hole
  symmetry, unlike the effective model for the spin model
  Eq.~\eqref{eq:spin-dimer}. Throughout this paper, we follow a notation
  similar to the one adopted in our previous work Ref.~\onlinecite{picon:08},
  but omit the tildes appearing therein. That is, for instance, amplitudes for
  the correlated processes for {\em holes} depicted in Fig.~\ref{fig:leapfrog}
  are here written as $s_{\rm 1,2}$ while in Ref.~\onlinecite{picon:08} those
  were denoted by $\tilde{s}_{\rm 1,2}$.}

\bibitem[{\citenamefont{Scalettar et~al.}(1995)\citenamefont{Scalettar,
  Batrouni, Kampf, and Zimanyi}}]{scalettar:95}
\bibinfo{author}{\bibfnamefont{R.~T.} \bibnamefont{Scalettar}},
  \bibinfo{author}{\bibfnamefont{G.~G.} \bibnamefont{Batrouni}},
  \bibinfo{author}{\bibfnamefont{A.~P.} \bibnamefont{Kampf}}, \bibnamefont{and}
  \bibinfo{author}{\bibfnamefont{G.~T.} \bibnamefont{Zimanyi}},
  \bibinfo{journal}{Phys. Rev. B} \textbf{\bibinfo{volume}{51}},
  \bibinfo{pages}{8467} (\bibinfo{year}{1995});

\bibitem[{\citenamefont{{Batrouni} and {Scalettar}}(2000)}]{batrouni:00}
\bibinfo{author}{\bibfnamefont{G.~G.} \bibnamefont{{Batrouni}}}
  \bibnamefont{and} \bibinfo{author}{\bibfnamefont{R.~T.}
  \bibnamefont{{Scalettar}}}, \bibinfo{journal}{Phys. Rev. Lett.}
  \textbf{\bibinfo{volume}{84}}, \bibinfo{pages}{1599} (\bibinfo{year}{2000}).

\bibitem[{\citenamefont{{Sengupta} et~al.}(2005)\citenamefont{{Sengupta},
  {Pryadko}, {Alet}, {Troyer}, and {Schmid}}}]{sengupta:05}
\bibinfo{author}{\bibfnamefont{P.}~\bibnamefont{{Sengupta}}},
  \bibinfo{author}{\bibfnamefont{L.~P.} \bibnamefont{{Pryadko}}},
  \bibinfo{author}{\bibfnamefont{F.}~\bibnamefont{{Alet}}},
  \bibinfo{author}{\bibfnamefont{M.}~\bibnamefont{{Troyer}}}, \bibnamefont{and}
  \bibinfo{author}{\bibfnamefont{G.}~\bibnamefont{{Schmid}}},
  \bibinfo{journal}{Phys. Rev. Lett.} \textbf{\bibinfo{volume}{94}},
  \bibinfo{pages}{207202} (\bibinfo{year}{2005}).

\bibitem[{\citenamefont{Schmidt et~al.}(2008)\citenamefont{Schmidt, Dorier,
  L\"auchli, and Mila}}]{schmidt:08}
\bibinfo{author}{\bibfnamefont{K.~P.} \bibnamefont{Schmidt}},
  \bibinfo{author}{\bibfnamefont{J.}~\bibnamefont{Dorier}},
  \bibinfo{author}{\bibfnamefont{A.~M.} \bibnamefont{L\"auchli}},
  \bibnamefont{and} \bibinfo{author}{\bibfnamefont{F.}~\bibnamefont{Mila}},
  \bibinfo{journal}{Phys. Rev. Lett.} \textbf{\bibinfo{volume}{100}},
  \bibinfo{pages}{090401} (\bibinfo{year}{2008}).

\bibitem[{\citenamefont{{Wessel} and {Troyer}}(2005)}]{SST}
\bibinfo{author}{\bibfnamefont{S.}~\bibnamefont{{Wessel}}} \bibnamefont{and}
  \bibinfo{author}{\bibfnamefont{M.}~\bibnamefont{{Troyer}}},
  \bibinfo{journal}{Phys. Rev. Lett.} \textbf{\bibinfo{volume}{95}},
  \bibinfo{pages}{127205} (\bibinfo{year}{2005});
%
\bibinfo{author}{\bibfnamefont{R.~G.} \bibnamefont{{Melko}}},
  \bibinfo{author}{\bibfnamefont{A.}~\bibnamefont{{Paramekanti}}},
  \bibinfo{author}{\bibfnamefont{A.~A.} \bibnamefont{{Burkov}}},
  \bibinfo{author}{\bibfnamefont{A.}~\bibnamefont{{Vishwanath}}},
  \bibinfo{author}{\bibfnamefont{D.~N.} \bibnamefont{{Sheng}}},
  \bibnamefont{and}
  \bibinfo{author}{\bibfnamefont{L.}~\bibnamefont{{Balents}}},
  \bibinfo{journal}{Phys. Rev. Lett.} \textbf{\bibinfo{volume}{95}},
  \bibinfo{pages}{127207} (\bibinfo{year}{2005});
%
\bibinfo{author}{\bibfnamefont{D.}~\bibnamefont{{Heidarian}}} \bibnamefont{and}
  \bibinfo{author}{\bibfnamefont{K.}~\bibnamefont{{Damle}}},
  \bibinfo{journal}{Phys. Rev. Lett.} \textbf{\bibinfo{volume}{95}},
  \bibinfo{pages}{127206} (\bibinfo{year}{2005});
  M. Boninsegni and N. Prokof'ev Phys. Rev. Lett. {\bf 95}, 237204 (2005).

\bibitem[{\citenamefont{{Abendschein} and {Capponi}}(2007)}]{abendschein:07}
\bibinfo{author}{\bibfnamefont{A.}~\bibnamefont{{Abendschein}}}
  \bibnamefont{and}
  \bibinfo{author}{\bibfnamefont{S.}~\bibnamefont{{Capponi}}},
  \bibinfo{journal}{Phys. Rev. B} \textbf{\bibinfo{volume}{76}},
  \bibinfo{pages}{064413} (\bibinfo{year}{2007}).

\bibitem[{rem()}]{remark_range2}
\bibinfo{note}{In this way, and by imposing that each cluster's low-energy
  spectrum is exactly reproduced, effective couplings of up to range-2 (see
  Fig.~5 in Ref.~\onlinecite{picon:08}) are computed.}

\bibitem[{\citenamefont{{Zhao} and {Paramekanti}}(2007)}]{zhao:07}
\bibinfo{author}{\bibfnamefont{E.}~\bibnamefont{{Zhao}}} \bibnamefont{and}
  \bibinfo{author}{\bibfnamefont{A.}~\bibnamefont{{Paramekanti}}},
  \bibinfo{journal}{Phys. Rev. B} \textbf{\bibinfo{volume}{76}},
  \bibinfo{pages}{195101} (\bibinfo{year}{2007}).

\bibitem[{\citenamefont{{Hassan} et~al.}(2007)\citenamefont{{Hassan}, {de
  Medici}, and {Tremblay}}}]{hassan:07}
\bibinfo{author}{\bibfnamefont{S.~R.} \bibnamefont{{Hassan}}},
  \bibinfo{author}{\bibfnamefont{L.}~\bibnamefont{{de Medici}}},
  \bibnamefont{and} \bibinfo{author}{\bibfnamefont{A.-M.~S.}
  \bibnamefont{{Tremblay}}}, \bibinfo{journal}{Phys. Rev. B}
  \textbf{\bibinfo{volume}{76}}, \bibinfo{pages}{144420}
  (\bibinfo{year}{2007}).

\bibitem[{\citenamefont{Murthy et~al.}(1997)\citenamefont{Murthy, Arovas, and
  Auerbach}}]{murthy:97}
\bibinfo{author}{\bibfnamefont{G.}~\bibnamefont{Murthy}},
  \bibinfo{author}{\bibfnamefont{D.}~\bibnamefont{Arovas}}, \bibnamefont{and}
  \bibinfo{author}{\bibfnamefont{A.}~\bibnamefont{Auerbach}},
  \bibinfo{journal}{Phys. Rev. B} \textbf{\bibinfo{volume}{55}},
  \bibinfo{pages}{3104} (\bibinfo{year}{1997}).

\bibitem[{agr()}]{agreement}
\bibinfo{note}{Similarly good agreement is found for the other couplings
  considered in Ref.~\onlinecite{chen:10}, $(J, J_{\times})=(0.38,0.23)$ and
  $(0.38,0.27)$.}

\bibitem[{1st()}]{1st_order}
\bibinfo{note}{A SS phase is obtained from the CORE-SCMFT procedure for {\em
  arbitrarily small} amplitudes for the leapfrog processes depicted in
  Fig.~\ref{fig:leapfrog}(a-b). First order transitions are only observed when
  correlated processes {\em vanish}, as it is the case for the higher-field
  transition out from the CBS plateau in Fig.~\ref{fig:MF}(b) (correlated
  hoppings for triplons vanish for all $J, J_{\times} \in [0,0.5]$).}

\end{thebibliography}
\end{document}